\documentclass{emulateapj}
\newcommand\ea{et al.\ }
\newcommand\hst{{\it HST}}
\def\chandra{{\it Chandra}}
\newcommand\psc{\ifmmode{\rm\,cm^{-2}}\else{${\rm\,cm^{-2}}$}\fi}
\submitted{To appear in ApJ Vol. 602, Feb. 10 2004}
\shorttitle{Accretion and Outflow in NGC 5135}
\shortauthors{Levenson et al.}

\begin{document}
\title{Accretion and Outflow in the AGN and Starburst of NGC 5135}

\author{N. A. Levenson\altaffilmark{1}, K. A. Weaver\altaffilmark{2,3}, T. M. Heckman\altaffilmark{3}, H. Awaki\altaffilmark{4}, and Y. Terashima\altaffilmark{5}}
\altaffiltext{1}{Department of Physics and Astronomy, University of Kentucky,
Lexington, KY 40506; levenson@pa.uky.edu}
\altaffiltext{2}{Code 662, NASA/GSFC, Greenbelt, MD 20771}
\altaffiltext{3}{Department of Physics and Astronomy, Bloomberg Center, Johns Hopkins University, Baltimore, MD 21218}
\altaffiltext{4}{Department of Physics, Faculty of Science, Ehime University,
Bunkyo-cho, Matsuyama, Ehime 790-8577, Japan}
\altaffiltext{5}{Institute of Space and Astronautical Science, 3-1-1 Yoshinodai, 
Sagamihara, Kanagawa 229-8510, Japan}

\begin{abstract}
Observations of the Seyfert 2 and starburst galaxy NGC 5135 with the {\it Chandra
X-ray Observatory} demonstrate that both of these phenomena contribute 
significantly to its X-ray emission.  We spatially isolate the active
galactic nucleus (AGN) and demonstrate that it is entirely obscured
by column density $N_H > 10^{24} \psc$, detectable
in the \chandra{} bandpass only as a strongly reprocessed, weak continuum
and a prominent iron K$\alpha$ emission line with equivalent width 
of 2.4 keV.
Most of the soft X-ray emission, both near the AGN and extending
over several-kpc spatial scales, is collisionally-excited plasma.
We attribute this thermal emission to stellar processes.
The AGN dominates the X-ray emission only at energies above 4 keV.
In the spectral energy distribution that extends to far-infrared
wavelengths,  nearly all of the emergent
luminosity below 10 keV is attributable to star formation, not the AGN.

\end{abstract}
\keywords{Galaxies: individual (NGC 5135) --- galaxies: Seyfert --- X-rays: galaxies}

\section{Introduction}

Active galactic nuclei and star formation are fundamentally related.
In their quiescent states, black hole and stellar components
are observed to be related in individual 
galaxies in the rough proportionality
between the black hole mass and the velocity dispersion of the
stellar spheroid \citep{Geb00,Fer00}, suggesting
a connection over cosmological timescales.
Also, \citet{Kau03} find a strong connection between active galactic nucleus (AGN) luminosity and the
age of the stellar population of the host galaxy for AGNs in the Sloan
Digital Sky Survey.
Locally,  AGNs and  strong 
star formation are  correlated, with roughly
half of well-selected samples of Seyfert 2 galaxies
exhibiting compact  circumnuclear starbursts
\citep[and references therein]{Cid01}. 
These Seyfert 2/starburst composite galaxies contain genuine AGNs, which
accretion onto the central black hole powers,
but the stellar processes can also become
energetically significant.

The local composite galaxies afford detailed examination
for discrimination of 
their AGN and stellar contributions and accurate measurement
of these combined effects.
The detailed studies serve as an important foundation for
understanding ultraluminous infrared galaxies, whose
bolometric luminosities exceed $10^{12} L_\sun$ \citep{San96}.
Among this class, the problem  is to 
identify  either an AGN
or star formation as
the primary energy source.  If present, an AGN may be
deeply buried, obscured both on small scales associated
directly with the central engine and on the large scales
of star formation \citep{Pta03}.
While the composite galaxies are not the direct
analogues of these complex higher-luminosity cases, they serve
as basic building blocks, demonstrating the primary
consequences of combining starbursts with AGNs in highly
obscured circumstances. 

Distinguishing these  X-ray emission processes is also important in the
study of the cosmic X-ray background (XRB).  
AGN can fundamentally produce the XRB \citep[e.g., ][]{Gia01}, but
they require a distribution of redshift and absorption
to match its observed spectrum
\citep[e.g., ][]{Set89}.
The outstanding difficulty is to identify the highly-obscured
populations, which are not always evident at energies below 10 keV.  
The Seyfert 2/starburst composite galaxies  cannot directly
account for the XRB, but they tend to be
strongly obscured  
\citep*{LWH}, 
and they illustrate
the systematic effects of observing AGNs in the
presence of starbursts, even at X-ray energies.

With high spatial resolution and simultaneous
spectroscopic data from the {\it Chandra X-ray Observatory},
we analyze here the X-ray emission due to
both AGN and starburst components of the Seyfert 2/starburst
composite galaxy NGC 5135.
This galaxy 
is optically classified as a 
Seyfert 2 on the basis of emission line
ratios \citep{Phi83}.
It also contains a powerful starburst
within 200 pc of the nucleus that is most evident
at ultraviolet energies \citep{Gon98}.
NGC 5135 is relatively nearby, at distance of 59 Mpc
(assuming $H_0=70 {\rm \, km\,s^{-1}\,Mpc^{-1}}$), 
so $1\arcsec$ corresponds to 285 pc.
In X-rays, we expect to find the characteristic unresolved
non-thermal continuum from accretion onto the central black hole
\citep{Tur97}.
We also distinguish the dominant X-ray signature of a starburst:
spatially extended thermal emission, due to 
individual supernovae and stellar winds, which collectively may
produce a ``superwind'' outflow that escapes the galaxy
\citep*{Dah98}. 
With the simultaneous spatial and spectral distinction of these
emission processes, 
we can more accurately measure them independently.
Relating the
X-ray data to multi-wavelength observations of this galaxy as a whole,
we draw general conclusions about identifying 
multiple energy sources in low-resolution observations of
galaxies that contain AGNs.

\section{Observations and Data Reduction}

The \chandra{} Advanced CCD Imaging Spectrometer (ACIS)
observed NGC 5135 on the back-illuminated S3 detector
on 2000 September 4.
(See \citealt{Wei00} 
for more information on \chandra{}.) 
We reprocessed all data from original Level 1 event files 
removing the spectral and  spatial randomization that is included in standard
processing with \chandra{} Interactive Analysis of Observations (CIAO)
software. 
We applied the current gain corrections for the S3 CCD (from 
2001 July 31), and 
we included only good events that 
do not lie on node boundaries, where discrimination of cosmic rays
is difficult.
We examined the lightcurves of background regions and found no
significant flares, so we used all data within the standard 
good-time intervals, for a total exposure of 29.3 ks.

\section{Image Analysis}
\begin{figure}[htb]
\includegraphics[width=0.5\textwidth]{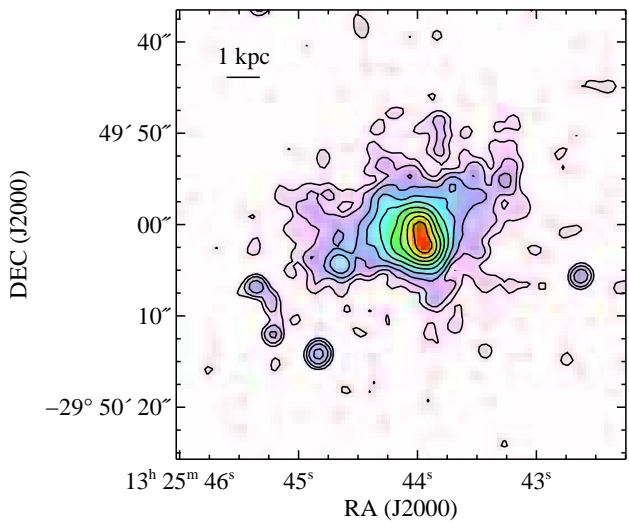} 
\caption{\label{fig:totimg}
\chandra{} X-ray image of NGC 5135.  
The broad-band (0.4--8 keV) image has been smoothed by 
a Gaussian of FWHM=$1\farcs5$.  The image is scaled
logarithmically.  
The overlaid intensity contours
begin $6\sigma$ above the background level and increase
by factors of two.  The AGN is
the northern central source, while the starburst is responsible
for the southern nuclear concentration
and diffuse emission.  Several additional  sources
are also apparent.
}
\end{figure}

\begin{figure}[hbt]
\includegraphics[width=0.5\textwidth]{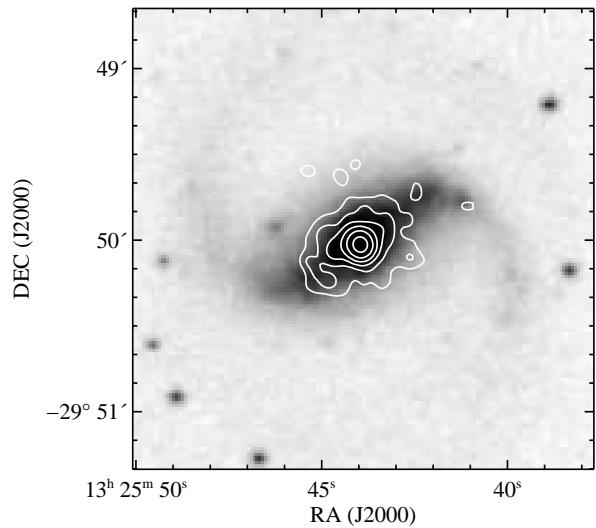} 
\caption{\label{fig:smoothxrayopt}
Digitized Sky Survey red image with 0.4--8.0 keV X-ray contours
overlaid.  The X-ray data have been smoothed by
a Gaussian of FWHM=$5\arcsec$ to illustrate the 
correlation of X-ray and  optical emission
on large scales.  The optical image is scaled
linearly, and the X-ray contours are scaled logarithmically.
}
\end{figure}

These \chandra{} data reveal the complexity of NGC 5135's X-ray emission,
as we illustrate with the 
total 0.4--8.0 keV image (Figure \ref{fig:totimg}).
The raw ($0\farcs5$) pixels have been smoothed by a Gaussian
of FWHM=$1\farcs5$, and both the image intensity and contours
are scaled logarithmically.
The active nucleus is not the only source of X-rays.  
Rather, the image reveals diffuse emission, off-nuclear point sources,
and, most surprisingly, {\em two} bright central sources,
separated by $2\arcsec$.  
As we demonstrate below, only the northern of these two 
is the active nucleus.
The highly-smoothed version of the image 
(Figure \ref{fig:smoothxrayopt}), shows that
the soft emission extends over a scale of approximately 
$50\arcsec \times 30 \arcsec$ (14 kpc $\times$ 9 kpc),
tracing the large-scale bar in the disk of NGC 5135.

\begin{figure}[thb]
\includegraphics[width=0.5\textwidth]{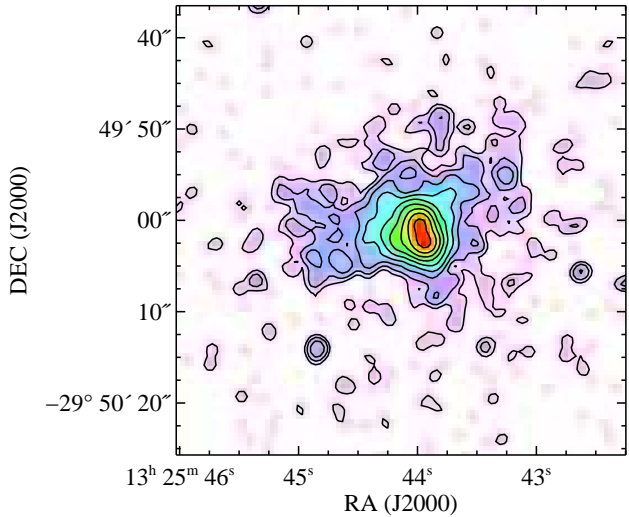} 
\caption{\label{fig:softimg}
\chandra{} soft X-ray (0.4--1 keV) image of NGC 5135,
smoothed by 
a Gaussian of FWHM=$1\farcs5$.  The image is scaled
logarithmically.  
The overlaid intensity contours
begin $8\sigma$ above the background level and increase
by factors of two. 
}
\end{figure}

\begin{figure}[thb]
\includegraphics[width=0.5\textwidth]{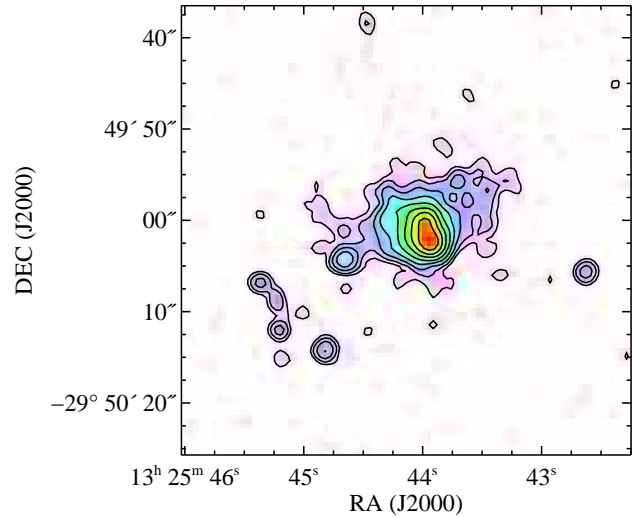} 
\caption{\label{fig:medimg}
\chandra{} medium X-ray (1--4 keV) image of NGC 5135,
smoothed by 
a Gaussian of FWHM=$1\farcs5$.  The image is scaled
logarithmically.  
The overlaid intensity contours
begin $6\sigma$ above the background level and increase
by factors of two. 
}
\end{figure}

\begin{figure}[htb]
\includegraphics[width=0.5\textwidth]{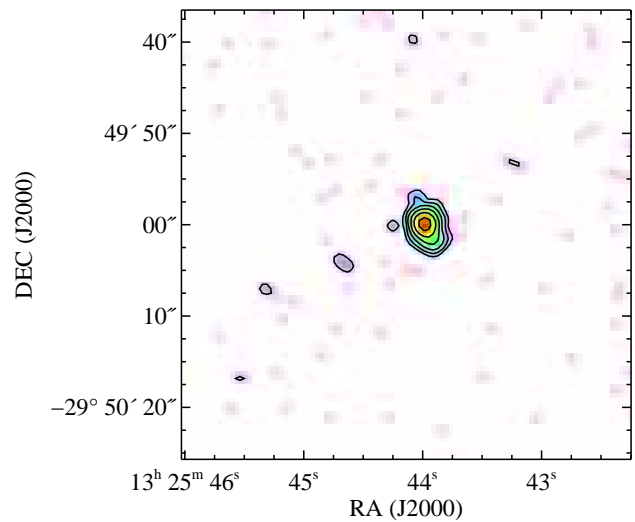} 
\caption{\label{fig:hardimg}
\chandra{} hard X-ray (4--8 keV) image of NGC 5135,
smoothed by 
a Gaussian of FWHM=$1\farcs5$.  The image is scaled
logarithmically.  
The overlaid intensity contours
begin $8\sigma$ above the background level and increase
by factors of two. 
The AGN is the strongest source, although resolved emission
due to the southern starburst concentration is also evident.
}
\end{figure}

We consider three broad energy regimes separately:
0.4--1.0, 1.0--4.0, and 4.0--8.0 keV, which we refer to as
the soft, medium, and hard images.
In soft X-rays, (Figure \ref{fig:softimg}), 
the galaxy appears  very similar to the total image.
The very extended emission is soft, and both central sources are
evident.  Some of the additional sources toward the east are
less prominent or do not appear at lower energies.
Some of the extended emission is still detected in the medium X-ray
(1--4 keV; Figure \ref{fig:medimg}) image, and the additional
sources are more prominent. In hard X-rays 
(4--8 keV; Figure \ref{fig:hardimg}),
the northern central source is strongest.  
The southern source contributes weakly, but is still evident, and two
of the eastern sources are significant.

We compare the two central sources with \chandra's point-spread function
(PSF) in the three separate energy bands.   In each of these three bands, we 
model the emission profile as a constant plus two two-dimensional
Gaussians having circular cross-sections.   
In all cases, the Gaussians are significantly broader than
a similar two-dimensional fit to the PSF at the appropriate energy
and detector position modeled with
the CIAO task mkpsf.
We conclude that the two central sources are 
resolved in all three broad bands.
Therefore, neither of the broad-band emission peaks is 
due to the unresolvable AGN alone.

We do, however, morphologically identify the AGN in the 
Fe K$\alpha$ emission of NGC 5135.  
This fluorescent line radiation arises near the
strong AGN continuum and
advantageously avoids
confusion with the soft, thermal starburst.
In this case, the Fe line is prominent above the continuum that either the
AGN or other sources could produce.
Isolating the emission in the 5.5--7.5 keV bandpass
and subtracting the continuum measured from 4 to 5.5 keV produces
an image in Fe K$\alpha$.
This image contains an unresolved point source, which
is located $0\farcs5$ 
north of the northern emission peak that is
measured in the total band image.
Thus, the Fe K$\alpha$ image 
specifically locates the AGN on the northern side of
the broad peak that appears in the total band image,
with both the AGN and other sources contributing to the
spatially-extended broad-band peak.
At the distance of NGC 5135, the point-like morphology 
constrains the physical scale of the Fe K$\alpha$ emission to
$R \lesssim 285$ pc.

\section{Spectroscopy}

We extracted spectra from several interesting regions.
In all cases, we also measured the local background
in nearby source-free regions and subtracted it.
Over the course of the \chandra{} mission, 
the soft X-ray sensitivity has diminished, likely the
result of build-up of material on the detector.
We have used the
``ACISABS'' model of G. Chatras and 
K. Getman\footnote{http://www.astro.psu.edu/users/chartas/xcontdir/xcont.html}
to create ancillary response files that account for
this time-varying effect.
In general, 
we grouped the spectra  into bins of a minimum of 30 counts,
so $\chi^2$ statistics are appropriate in the model fitting,
which we performed in XSPEC \citep{Arn96}.
In the best-fitting models discussed below,
only data from 0.4 to 8.0 keV are considered, and 
the inclusion of additional model components and their
free parameters are significant at a minimum of the 95\% confidence
limit, based on an $F$ test.  All quoted errors are 90\% confidence
for one parameter of interest.

Observational results from ``pure'' AGN and ``pure'' starbursts
physically motivate the spectral models we consider, and we
recognize that all components may be present here, even in
spatially-restricted regions.
The AGN produces an X-ray continuum that is 
characterized by a power law of photon index $\Gamma \approx 1.9$,
as observed in Seyfert 1 galaxies \citep{Nan94}, assuming the
central engines of Seyfert 1s and Seyfert 2s are intrinsically identical.  
With modest column density along the line of sight, the
soft X-rays will be photoelectrically absorbed.  When the obscuration
becomes Compton thick, which is the case here, the emergent continuum
is entirely reflected and is strongly reprocessed, to
$\Gamma \approx 0$ \citep*[e.g., ][]{Kro94} within the observed energy
regime.
Fluorescent Fe emission is common in the X-ray spectra of
Seyfert galaxies.  The Fe K$\alpha$ line 
appears at  central energy $E_c = 6.4$ keV
in material
less ionized than \ion{Fe}{17}.
If the iron is nearly fully ionized,
 $E_c = 6.7$ or 6.9 keV, for
He-like or H-like ions, respectively. 
Each of these 
fluorescence lines may be spectrally 
characterized by a single Gaussian.
The starburst produces X-rays in both point sources
and genuinely diffuse gas.  
Observed starburst spectra tend to be soft, indicating multiple
thermal components in the diffuse material as well as a hard continuum
\citep[e.g., ][]{Dah98,Dah00}.  The
X-ray binaries produce the power-law continuum, although the integrated
spectrum of the unresolved sources may not be characteristic
of any individual sources. 

\tabletypesize{\tiny}
\begin{deluxetable*}{llllllllllllc}
\tablewidth{0pt}
\tablecaption{Spectral Model Paramters\label{tab:modelpars}}
\tablehead{
\colhead{Region}
&\colhead{$N$\tablenotemark{a}$_H$}
&\colhead{$kT$\tablenotemark{b}$_1$}
&\colhead{$A$\tablenotemark{c}$_1$}
&\colhead{$kT$\tablenotemark{b}$_2$}
&\colhead{$A$\tablenotemark{c}$_2$}
&\colhead{$\Gamma$}
&\colhead{$A$\tablenotemark{d}$_3$}
&\colhead{$E$\tablenotemark{e}$_{line}$}
&\colhead{$EW$\tablenotemark{f}$_{line}$}
&\colhead{$F$\tablenotemark{g}$_{0.5-2}$}
&\colhead{$F$\tablenotemark{h}$_{2-10}$}
&\colhead{$\chi^2/$dof}
}
\startdata
North\tablenotemark{i} & $42^{+33}_{-21}$ & $0.05^{+0.02}_{-0.01}$ & $(3.4^{+24}_{-2.1})\times 10^4 $  & $0.57\pm0.09$  & $3.8^{+0.9}_{-0.6}$  &
$0.0^{+1.6}_{-0.2}$  & $0.21^{+0.04}_{-0.18}$   &  $6.39^{+0.03}_{-0.04}$ &   $2.4^{+1.8}_{-0.5}$  &
 $6.1^{+1.5}_{-1.0}$ & $22\pm4.4$ & 99/93 \\
& \nodata & \nodata & \nodata & \nodata & \nodata & \nodata & \nodata & $1.77\pm 0.05$ & $0.26^{+0.13}_{-0.15}$ & \nodata & \nodata & \nodata \\
North\tablenotemark{j} & $21^{+1.2}_{-1.3}$ & $0.06\pm 0.003$ & 
$(1.7^{+0.3}_{-0.2})\times 10^3$ & $0.58^{+0.07}_{-0.09}$ &
$2.7^{+0.2}_{-0.6}$ & $0.3^{+1.3}_{-0.1}$ & $0.38^{+0.8}_{-0.1}$
& $6.39^{+0.03}_{-0.04}$ & $3.0^{+1.4}_{-0.8}$ & $6.8^{+0.5}_{-1.4}$ &
$21^{+2.9}_{-6.8}$ & 442/509 \\ 
& \nodata & \nodata & \nodata & \nodata & \nodata & \nodata & \nodata & 
$1.77^{+0.06}_{-0.04}$  
 & $0.17^{+0.15}_{-0.09}$ & \nodata & \nodata & \nodata \\
South & $ 11^{+5.7}_{-5.0}$ & $0.66\pm 0.05$   & $3.8^{+0.8}_{-0.7}$ & \nodata & \nodata   & $2.6^{+0.4}_{-0.3}$  & $2.9^{+1.1}_{-0.8}$  &\nodata&\nodata & $11\pm 2$ & $3.6^{+1.3}_{-0.9}$ & 39/35 \\
D1   &$8.4^{+7.1}_{-3.7:}$ & $0.62\pm0.07$  & $2.3^{+0.8}_{-0.6}$ & \nodata & \nodata & $2.2^{+0.5}_{-0.4}$  & $1.8\pm0.6$   &\nodata&\nodata & $7.4^{+2.5}_{-1.9}$ & $3.7^{+1.0}_{-1.1}$ & 39/27 \\
D2   & $42^{+15}_{-26}$ & $0.25^{+0.14}_{-0.06}$ & $14^{+66}_{-4.7}$ &\nodata&\nodata& 2.2f  & $0.63\pm 0.3$  &\nodata&\nodata & $3.6^{+2.0}_{-0.9}$ & $1.2\pm0.5$ & 21/13 \\
\enddata

\tablenotetext{a}{Column density in units of $10^{20}{\rm\,cm^{-2}}$.}
\tablenotetext{b}{Temperature of thermal plasma in keV.}
\tablenotetext{c}{Normalization of thermal component in units of $10^{-5} K$, 
where $K=(10^{-14}/(4\pi D^2))\int n_e n_H dV, D$ is the distance to the source (cm), 
$n_e$ is the electron density (${\rm cm^{-3}}$), and $n_H$ is the hydrogen density 
(${\rm cm^{-3}}$).}
\tablenotetext{d}{Normalization of power law in units of
$10^{-5} {\rm\,photons\, keV^{-1}\,cm^{-2}\,s^{-1}}$ at 1 keV.}
\tablenotetext{e}{Energy of line center in keV.}
\tablenotetext{f}{Equivalent width of line in keV.} 
\tablenotetext{g}{0.5--2.0 keV model flux in units of
$10^{-14}{\rm\,erg\,cm^{-2}\,s^{-1}}$.}
\tablenotetext{h}{2.0--10.0 keV model flux in units of
$10^{-14}{\rm\,erg\,cm^{-2}\,s^{-1}}$.}
\tablenotetext{i}{Parameters and errors on $\Gamma$, $A_3$, and high-energy components of $E_{line}$ and $EW_{line}$
are based on high-energy fit alone.}
\tablenotetext{j}{Parameters and errors are based on full spectral fit, using
the C-statistic.}
\tablecomments{
Errors are 90\% confidence limits for one interesting parameter.
Parameters that are constrained by hard limits are marked with a colon.
Fixed parameters are marked with f.}

\end{deluxetable*}
\tabletypesize{\footnotesize}

\subsection{The AGN Region}
\begin{figure}[hbt]
\includegraphics[angle=270,width=0.5\textwidth]{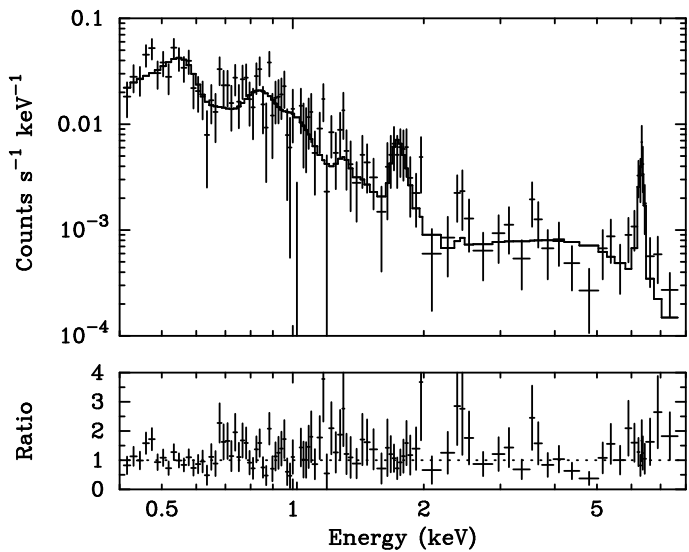} 
\caption{\label{fig:specn}
\chandra{} ACIS-S3 spectrum of  the northern central source.
This region includes the NGC 5135 AGN, whose direct
X-ray emission is completely obscured.
A complex model consisting of two thermal components,
a power law, and two emission lines
fits the data well.
In the upper panel, 
the data are plotted as crosses, and the model, convolved with
the detector response, is plotted as a histogram.
The ratio of data/model is plotted in the lower panel.
}
\end{figure}

We centered the extraction of the northern central source
on the Fe K$\alpha$ center measured in the images.
We measured the spectrum in a region of only $1\farcs2$ radius
to avoid contamination by the southern source.
The best-fitting model requires two thermal components and a Gaussian 
at soft energies 
in addition
to  a flat power law continuum and a Gaussian line, which are
characteristic of the AGN at higher energies.
We apply the same absorption by a foreground screen 
to all of these components.
Figure \ref{fig:specn} contains 
the data, model, and ratio of data/model, and Table \ref{tab:modelpars} lists
the model parameters.
We had previously fit the 4--8 keV spectrum \citep{Lev02}
with a power law and a single Gaussian 
to measure the Fe K$\alpha$ line and its underlying continuum.
In order to retain significant information at energies above the
line, we fit the unbinned spectrum using the C-statistic \citep{Cas79}.
The Fe line is spectrally unresolved, with equivalent width
 EW $= 2.4$ keV.  The power law is flat ($\Gamma = 0$), characteristic
of a pure reflection spectrum.  
The large EW and flat spectral slope demonstrate that
the AGN is not viewed at all directly within the \chandra{} bandpass;
it must be obscured by $N_H > 10^{24} \psc$.

These results most accurately describe the high-energy emission, so
we fix these parameters while fitting the entire 0.4--8 keV binned
spectrum.  In the complete spectrum, two additional components are
necessary to account for most of the soft X-ray emission.
We use the thermal equilibrium  model of 
Mewe and Kaastra \citep{Mew85,Arn85,Mew86}, with updated Fe L
calculations \citep{Lie92} and solar abundance.
The two thermal components are at temperature $kT = 0.05$ and $0.6$ keV.
We note that the normalization of the softer component is several
 orders of magnitude
greater than that of the harder component.  The normalization 
is proportional to the volume integral of the square of the density, and a greatly
increased density in the low-temperature region is the likely origin
of this difference.

Alternatively, we  fit the entire 0.4--8 keV spectrum simultaneously.
Again, we use the unbinned data and the C-statistic to retain some
information above the high-energy emission line.
These parameters are not significantly different from
the previous fit and are  listed in the second row of 
Table \ref{tab:modelpars}.
The slightly steeper photon index ($\Gamma = 0.3$) contributes to the observed 
1--2 keV emission,
and the high-energy line EW is still large (EW = 3.0 keV).

Despite the uncertainties in measuring $\Gamma$ and
its normalization independently, the Fe EW remains
large in both the high-energy and full-spectrum fits.
Considering the errors on
these parameters jointly, a steeper photon index does require
a higher normalization (measured at 1 keV), although the
magnitude of the high-energy continuum change is small.

The flat power law of the high-energy model is a consequence of reflection.
We also considered a more complete model based on Monte
Carlo simulations of Compton reflection by a cold
population of electrons \citep[the PEXRAV model in XSPEC]{Mag95}, 
which further demonstrates that we observe only
the reflected continuum and does not  improve the
fits at all.   Specifically, we adopted standard values of
the parameters to which we have no sensitivity,
 fixing the  high-energy cutoff $E_c = 100$ keV
and inclination angle $i = 63^\circ$.
We first considered a fixed 
intrinsic photon index $\Gamma = 2.0$.
Fitting the full spectrum 
we find the unphysical result that the 
magnitude of the reflected spectrum relative to the observed
intrinsic spectrum exceeds 30, and the Fe K$\alpha$ EW = 2.9 keV. 
The photon index cannot be reasonably constrained.
If it is free parameter, we find $\Gamma = 2\ (+3, -2)$.

With the spectral resolution and high-energy sensitivity
of the present data, there is no discernable difference
between a pure reflection model and the simplified flat power law 
that we have adopted to describe it.  
Above 4 keV, the only formal
difference between these two models is the Fe absorption
edge near 7 keV, which we cannot detect.  
The softer absorption edges of other elements in
the reflection model are
not strong enough to be detected where 
thermal emission dominates, despite
\chandra's improved sensitivity  at lower energies.

We considered the possibility that photoionization by the AGN might
produce the soft X-ray emission in
the northern region.  
We 
attempted to fit the spectrum with photoionization models
produced with XSTAR\footnote{http://heasarc.gsfc.nasa.gov/docs/software/xstar/xstar.html}
instead of the thermal
plasma components.
We also compared NGC 5135 with the observed photoionization
spectrum of NGC 1068 \citep{Kin02} because the models
fail to reproduce some of the line emission that is detected
in this Seyfert galaxy.  None of these photoionization 
spectra accurately reproduce the observed emission of NGC 5135.
In general, NGC 5135 exhibits several broad emission peaks that
are due to complexes of many unresolved lines, characteristic
of thermal emission.  Even at the relatively low spectral resolution
of these observations, the emission features of all the
photoionization spectra appear too sharp.

Finally, in addition to the Fe K line at 6.4 keV and the 
thermal emission, the northern source also
requires a second line at 1.8 keV.  
This line is consistent with neutral 
silicon fluorescence, both in central energy
and equivalent width \citep{Mat97} for the geometry that the Fe emission
constrains (\S \ref{subsec:fek}).  We searched for other fluorescent lines
that are expected to be relatively strong (EW $>$ 100 eV).
The emission measured near 2.4 keV is suggestive of neutral sulfur and
can be fit with an unresolved Gaussian of EW = 340 eV.  We caution, 
however, that this
component is not statistically significant in the model fit, and
the central energy is significantly  (1.8$\sigma$) higher than the
2.31 keV expected for sulfur.  Some of the lower-energy fluorescent
lines, including oxygen, neon, and magnesium, should have
large EWs with respect to the reflected continuum, but
we cannot discern them against the strong soft thermal emission.

\begin{figure}
\includegraphics[angle=270,width=0.5\textwidth]{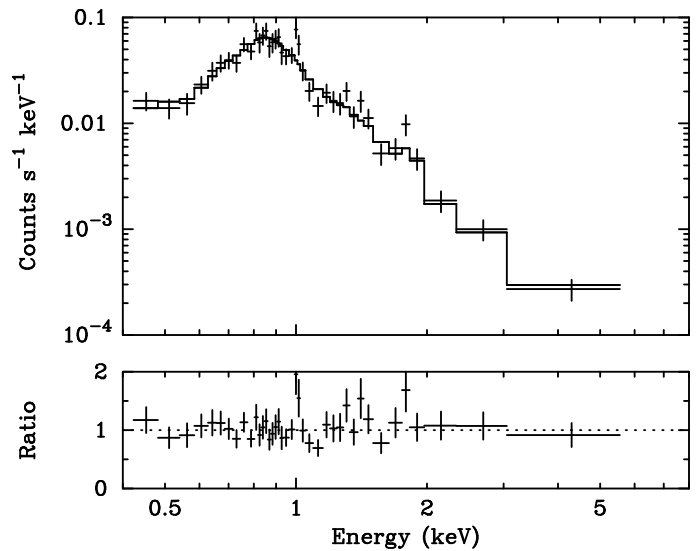} 
\caption{\label{fig:specs}
\chandra{} ACIS-S3 spectrum of  the southern central source,
the strong circumnuclear starburst of NGC 5135.
Typical of starbursts, the data require a thermal component and a hard
X-ray continuum.
}
\end{figure}

The prominent inner-shell fluorescence lines are 
characteristic of pure reflection from an optically thick medium.
Photoionization of optically thin, low density material
does produce line emission, as in
the X-ray spectrum of NGC 1068 \citep{Bri02,Kin02}.
In NGC 5135, however, an additional photoionized component would
also produce many other strong emission lines, which we do not observe.
As we note above, 
photoionization alone, {\em instead of} thermal emission, fails
to reproduce the soft X-ray spectrum. 
Alternatively, we consider pure collisional excitation,
which does produce some Si
line emission at similar energies, as observed in other
starburst galaxies  
 \citep{Wea02} and Compton thick AGNs \citep{Bol03}.
The line component we model here, however, is required {\em in addition to}
the contributions from the thermal plasmas that are also
present and dominate the soft X-ray emission.  
Thus, we conclude that fluorescent Si 
is the likely
origin of this additional emission line.

\subsection{The Compact Starburst}
We extracted the spectrum from a region of $1\farcs2$ radius
to isolate the southern central source, which
we associate with the strong circumnuclear starburst.
The X-ray emission is spatially resolved in all the broad bands
measured above.
The best-fitting model requires one thermal component and a 
power law, both of which are absorbed by the same column density. 
The temperature of the thermal component is 0.7 keV, similar to pure
starburst galaxies.  The thermal emission accounts for most of the
soft X-rays, and the  power law contributes all of
the X-ray emission above 4 keV.
The extinction along the line of sight, $N_H = 1.1\times10^{21}\psc$, is
greater than that due to the Milky Way Galaxy alone, although it is 
comparable to the value \citet{Gon98} measure ($N_H = 9.9\times10^{20}\psc$) 
at ultraviolet wavelengths in a similarly-sized aperture.

The non-thermal power law is likely the net effect of individual sources,
primarily X-ray binaries, in the starburst.
The slope of the power law is very steep: $\Gamma = 2.6$.  Although
this is atypical of  any individual
X-ray binary system, it is characteristic of the integrated spectrum
of distinct sources measured in face-on star-forming galaxies, which
contain many observable sources \citep{Zez02}.
We considered fixing the photon index at a value more typical of
individual non-thermal X-ray  sources ($\Gamma=1.9$), but found 
the model fit to the data significantly worse, even allowing for
additional soft X-ray components.
Figure \ref{fig:specs} shows this compact starburst 
spectrum with the best-fitting
model and ratio of data/model, and 
Table \ref{tab:modelpars} lists the best-fitting model parameters.

\subsubsection{Abundance Variations}
We also considered thermal plasmas with variable abundances.
A starburst wind may be enriched if it carries the
metals that supernovae produce.  On the other hand, 
the wind may have low abundances, either because it 
consists of unenriched halo material or because
refractory elements are not present in the gas phase.
Low abundances have been reported in X-ray observations
of starburst galaxies, but these results may be due to
the systematic difficulties of measuring multi-temperature
plasmas or neglecting underlying continuum contributions
\citep{Wea00,Str02}.  

We find slight evidence for non-solar
abundances in the southern region of NGC 5135.
Considering variable abundances in the thermal component of 
the best-fitting model above (which includes a non-thermal
continuum contribution), the iron abundance is best fit
at [Fe/H] = 0.39 (+0.14, -0.13) [Fe/H]$_\sun$, where
[Fe/H]$_\sun = 4.68\times10^{-5}$ \citep{And89}.
We obtain similar results if other refractory elements 
(Na, Al, Ca, and Ni) are constrained to have the same relative
abundance as Fe:  [Fe/H] = 0.37 (+0.17, -0.12) [Fe/H]$_\sun$.
The plasma temperature remains the same, although the
power law becomes somewhat flatter ($\Gamma = 2.0$). 
Starburst spectra often  show enhanced abundances of alpha elements
with respect to Fe \citep{Mar02,Str03}.
Although we cannot simultaneously constrain Fe and alpha abundance variations
independently in these data, we do observe this effect generally.
Above, varying only Fe abundance, $Z_\alpha/Z_{Fe} = 2.6$ (+1.3, -0.7).
Alternatively, fixing Fe at solar abundance and allowing
the common abundance of O, Ne, Mg, Si, and Ca to vary, 
$Z_\alpha/Z_{Fe} = 2.3$ (+1.1, -1.0), with  $\Gamma= 2.3$.
The abundance of oxygen most strongly determines the value of 
this ratio, and here we have 
adopted [O/H]$_\sun = 8.51\times10^{-4}$ \citep{And89}.

We also considered multi-temperature thermal plasmas with variable
abundances without any non-thermal component.  These fits were statistically
worse and physically unlikely.  Even with the abundance variations,
these fits require an extremely high 
temperature plasma ($kT \approx 10$ keV) to account for the emission
at energies above 4 keV.

\subsection{Diffuse Emission}
We extracted two regions of diffuse emission.  The first (D1) is
the brighter central region of maximum radius $6\farcs5$, excluding 
an ellipse that encompasses the extremely bright north and south
sources and a point source within the region of interest.
The second (D2) is elliptical, extending 
$14\arcsec \times 10\arcsec$ at position angle $PA = 118^\circ$,
excluding the central circle of D1 and interior point sources.

\begin{figure}[hbt]
\includegraphics[angle=270,width=0.5\textwidth]{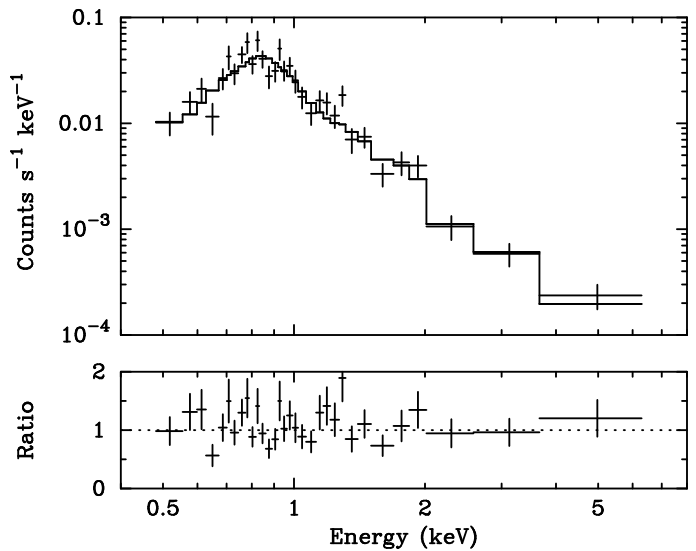} 
\caption{\label{fig:specx2}
\chandra{} ACIS-S3 spectrum of central diffuse emission (region D1).
A thermal component and a power law are required to fit the data.
}
\end{figure}

We fit D1 with an absorbed thermal component and 
a power law (Figure \ref{fig:specx2}, Table \ref{tab:modelpars}). 
For the thermal emission, $kT = 0.63$ keV,
and the power-law  photon index $\Gamma = 2.2$.
With the total column density the same for all components and
a free parameter, we find $N_H = 8.4\times 10^{20} \psc$ in 
the best-fitting model,
but it is not significantly different from
 the Galactic contribution alone, of  
$N_{H,MW} = 4.7\times 10^{20} \psc$.
The thermal plasma accounts for most of the soft X-rays, 
and the power law is responsible for nearly all the hard X-rays.
Spectrally, the extended central region is very similar to
the southern central source, although it has a much lower surface
brightness.  
The ratio of hard to soft luminosity is 
somewhat larger in D1, 
compared with the southern source.  In this case,
normal disk sources, in addition
to those in the areas of significant star formation, likely become
increasingly important.

We considered abundance variations in the D1 spectrum but found
no evidence for non-solar abundances.  
Similar to the results of the
South region, excluding the non-thermal component
cannot reproduce the higher-energy spectrum of D1,
even using multi-temperature, variable abundance models.

\begin{figure}[htb]
\includegraphics[angle=270,width=0.5\textwidth]{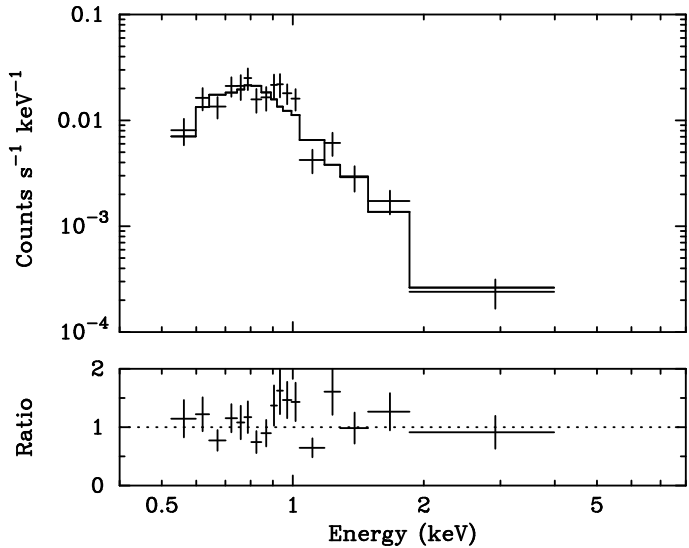} 
\caption{\label{fig:specxe}
\chandra{} ACIS-S3 spectrum of very extended diffuse emission (region D2).
A single thermal component and power law fit the data.
}
\end{figure}
The spectrum of the largest region, D2, also requires 
a single-temperature plasma.  This
component is much cooler ($kT = 0.25$ keV) and
more absorbed  ($N_H = 4.2\times 10^{21} \psc$).
The required second component is not well constrained.  We
model it as a power-law, fixing $\Gamma = 2.2$, the
best-fitting value of $\Gamma$ in the D1 spectrum.
The best-fitting model of D2 is plotted with the spectrum in
Figure \ref{fig:specxe}, and the model parameters are listed
in Table \ref{tab:modelpars}.

A multi-temperature medium may be a more
accurate description of physical conditions of this region,
possibly accounting for the unphysical conclusion that this region
is more obscured than D1.
With the limited sensitivity of these data, however, 
more complex models are not statistically significant.
Examining a model that includes a second thermal plasma component,  we
find $kT_2 \approx 0.9$ keV.  With this
higher-temperature component accounting for the 
$E \approx 1$ keV emission, the normalization of the very soft
component is lower.  The column density is also slightly lower 
($N_H = 3.1\times 10^{21} \psc$), although still larger than that 
measured in D1.  

\subsection{Additional Sources}

We detect several other weaker sources within 2$\arcmin$ of the center of NGC
5135, which   we list in Table \ref{tab:ptsrc} in order of distance from
the nucleus.  The last three are unlikely to reside within the
galaxy, but we include them for completeness, since they are located within a few
arcseconds of the optically-detected spiral arms.
These additional sources are interesting as possible 
intermediate-luminosity X-ray objects (IXOs), those
with X-ray luminosity $L_X > 10^{39} {\rm \, erg\,s^{-1}}$.
Exceeding the Eddington luminosity
of a stellar-mass black hole, 
they could  be single sources if they are 
more massive, accrete above the Eddington rate,
or emit anisotropically. 
Although IXOs are observed in galaxies without intense
ongoing star formation \citep{Col02},
in galaxies having high star formation rates, 
a significant population of IXOs 
is correlated with the star formation rate \citep{Col03}.
Not all of the sources listed in Table \ref{tab:ptsrc} are 
intermediate-mass ($10^2$--$10^5 M_\sun$) black hole candidates, however.
Four of the NGC 5135 sources
(2, 3, 5, and 6) are certainly spatially extended, and 
even the unresolved sources encompass spatial scales of several
hundred parsecs, so they could be clusters of multiple sources.

\begin{deluxetable*}{lrrrrrcc}
\tabletypesize{\footnotesize}
\tablewidth{0pt}
\tablecaption{Additional Sources\label{tab:ptsrc}}
\tablehead{
\colhead{Source}&
\colhead{R.A.}&\colhead{Decl.}
&\colhead{Counts}
&\colhead{Counts}
&\colhead{Counts}
&\colhead{$F_{0.3-8}$}
&\colhead{$L_{0.3-8}$}\\
&\colhead{(J2000)}&\colhead{(J2000)}
&\colhead{(0.3-8 keV)}
&\colhead{(0.3-2 keV)}
&\colhead{(2-8 keV)}
&\colhead{($10^{-15} {\rm \,erg\,cm^{-2}\,s^{-1}}$)}
&\colhead{($10^{39}{\rm \,erg\,s^{-1}}$)}
}
\startdata
1 & 13 25 43.7 & -29 49 56    &   46  &43 & 3    &  7.7     &    3.2 \\ 
2  & 13 25 44.7 & -29 50 04   &   41  &26 & 15   &  7.5     &    3.2 \\ 
3  & 13 25 44.8 & -29 50 14   &   25  &20 & 4    &  6.0     &    2.5 \\ 
4  & 13 25 42.6 & -29 50 06   &   13  &13 & 0    &  6.4     &    2.7 \\ 
5  & 13 25 45.3 & -29 50 08   &   20  &14 & 5    &  4.9     &    2.1 \\ 
6  & 13 25 45.2 & -29 50 12   &    8  &7  & 0    &  3.6     &    1.5 \\ 
7  & 13 25 44.1 & -29 49 33   &    8  &7  & 1    &  2.7     &    1.1 \\ 
8 & 13 25 45.3 & -29 49 35    &    6  &5  & 1    &  1.9     &    0.80 \\
9  & 13 25 48.0 & -29 49 48   &   35  &31 & 5    &  11      &    4.6 \\ 
10 & 13 25 46.7 & -29 51 43   &    9  &7  & 2    &  0.98    &    0.41 \\
11  & 13 25 44.1 & -29 48 10  &   68  &46 & 22   &  13      &    5.5 \\ 
12  & 13 25 52.8 & -29 50 09  &   30  &25 & 10   &  5.4     &    2.3 \\ 
\enddata

\tablecomments{Units of right ascension are hours, minutes, and
seconds, and units of declination are degrees, arcminutes, 
and arcseconds.}
\end{deluxetable*}

The net source counts in the total (0.3--8 keV), soft (0.3--2 keV), and
hard (2--8 keV) bands are listed in columns 4 through 6, respectively,
where the local background 
measured in a surrounding annulus has been subtracted.
We use these energy bands, rather than the 0.5--2 and 2--10 keV
that are more common in the literature and presented
elsewhere in this work to avoid errors
in the model-dependent conversion of the observed data
to unmeasured energies.
In some cases, the total number of 
background-subtracted counts detected
does not equal the sum of the soft- and hard-band measurements
because of rounding errors.
None of the sources is bright enough to distinguish 
the true physical character of its emission, but
soft and hard count rates indicate variations of spectral hardness
in the different sources.  We note that many of these sources
are extremely soft, unlike the harder emission 
typically measured in
IXOs. 
The emission of these soft sources
could therefore be thermal, due to concentrations of
supernovae and stellar outflows, although
we cannot rule out accretion, which produces soft
spectra in some IXOs \citep{Ter03}.

To estimate fluxes and luminosities consistently, we model each spectrum
as an absorbed power law, even though this may not be physically
appropriate.  We fix the photon index
$\Gamma = 1.8$, and allow absorption to vary 
in excess of the Galactic column density
along this line of sight.  The observed fluxes and luminosities 
(not corrected for absorption) in
the 0.3--8 keV band are listed in columns 7 and 8 of Table \ref{tab:ptsrc},
where the luminosity is calculated for the 59 Mpc distance to NGC 5135.

\section{Physical Interpretation}
\subsection{Identification of X-ray Emission Sources\label{subsec:origin}}
The X-ray spectral and spatial information together identify 
the northern source as the AGN of NGC 5135.
The most compelling evidence is the Fe K$\alpha$ line,
which stellar processes cannot produce with such a large
luminosity or equivalent width.  
The spectral aperture by necessity covers a relatively
large area, so in addition to the AGN,
it includes other emission sources.  Therefore,
the energy-restricted Fe-line image is required
to spatially constrain the AGN.  In this
uncontaminated image of the central engine, we demonstrate
that the AGN is unresolved.

At higher spatial resolution, 
we also identify the AGN located north of extended stellar
emission in {\it Hubble Space Telescope (HST)} optical and near-infrared (NIR)
images.
NGC 5135 was observed for 500s with the Wide-Field and Planetary Camera 2 (WFPC2)
through the F606W filter on 1995 February 15
and for 320s  with NICMOS Camera 2 through the F160W filter
on 1998 May 28.
In both the optical and NIR,
the nucleus
is located at the apparent center of relatively bright
emission that extends over scales of $5\arcsec$.
 In the latter, at 1.6$\mu$m, it is unresolved and the brightest source.
In the former, the nuclear emission is resolved, including a significant
contribution that
extends over physical scales greater than 10 pc.
The strongly-obscured AGN is not the brightest source observed within
the 606W filter bandpass.  
In the ultraviolet, the AGN is a 
weak source compared with nearby stellar emission,
as observed at 2150\AA\ 
with the F210M filter of {\it HST}'s Faint Object Camera
for 1963s on 1995 July 17.  (We note that the 
AGN is located approximately $0\farcs2$ west and $1\farcs7$ north of
the UV-brightest point and is {\em not} the location that
\citet{Gon98} tentatively identified as the nucleus.  In fact,
their UV spectral aperture probably excluded most of the AGN light.)

\begin{figure}[hbt]
\includegraphics[width=0.5\textwidth]{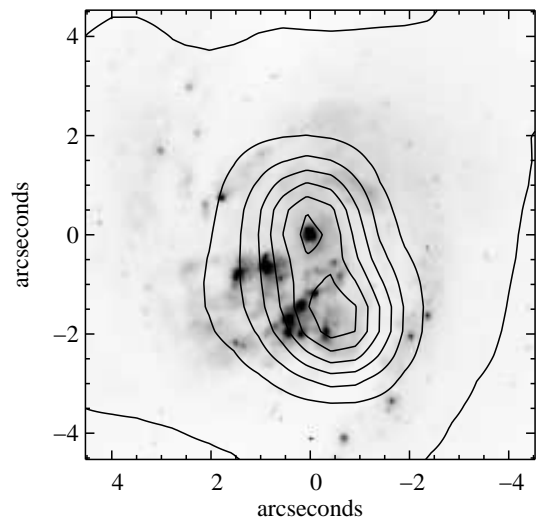} 
\caption{\label{fig:hst}
{\it HST} optical (6060\AA) image with \chandra{} total-band
contours overlaid, scaled linearly. North is up and east is to the left.
The AGN is located 
at the origin, with  the X-ray emission shifted to align
at the northern X-ray source.
In both images, the AGN is north of prominent, extended emission.
The high-resolution optical data also illustrate that stellar sources,
in addition to the AGN, are contained within the small nuclear X-ray aperture.
}
\end{figure}

Because of errors in the absolute astrometry of 
both \chandra{} and {\it HST}, 
the location of the northern X-ray source does not appear to
be exactly coincident with the UV, optical, or NIR AGN, based
on the telescopes' reported astrometry alone.  
In Figure \ref{fig:hst}, we illustrate the approximate relationship 
between the X-ray and optical emission, having shifted
the X-ray contours by $1\farcs8$ to align the northern source with the 
optical AGN and resampled onto a common pixel scale.  A further clockwise rotation of the X-ray
image may also be suggested, but the simple planar shift alone
illustrates the strong stellar contribution south of the AGN,
near the southern X-ray source.

The AGN is not the only source of X-ray emission within the
340-pc northern spectral aperture.
Mass is strongly concentrated in the nuclear region,
so it is a location of the most intense star formation.
At the higher resolution of WFPC2,
we distinguish many individual stellar sources at optical wavelengths
on sub-arcsecond scales.
Stellar processes, including supernovae, stellar winds, and 
large-scale outflows, are likely responsible for the soft thermal X-ray
emission, even in the immediate surroundings of the AGN.
Most of the thermal emission in the northern region is due to the
0.6-keV component, which is similar to the only significant thermal
components of the South and D1 regions.  Thus, we suggest
that even within the northern aperture, the thermal emission
is stellar, not due to the AGN.

At 6 cm, VLA observations show concentrated emission at
$\alpha = 13^{\rm h} 25^{\rm m} 43.^{\rm s}94$,
$\delta = -29^\circ 50\arcmin 02\farcs8$ (J2000),
with some low-level emission extending toward the northeast \citep{Ulv89}.
The radio core appears to be offset $2\farcs9$ south of the AGN, 
which is located in the
continuum-subtracted Fe K$\alpha$ image at
$\alpha = 13^{\rm h} 25^{\rm m} 43.^{\rm s}97$,
$\delta = -29^\circ 50\arcmin 00\farcs0$ (J2000).
The radio core appears aligned instead with the southern X-ray source
(within $0\farcs3$), suggesting that the concentrated radio emission
is due to the starburst.

The southern source is simply a starburst.
The thermal emission and the hard continuum
are typical spectral components measured in local starbursts.
The power-law continuum in the starburst region is therefore unrelated
to the nearby AGN in this galaxy.
Although only one thermal component is significant in this region,
the reality is likely more
complicated, with a range of temperatures actually present \citep{Str00}.

The starburst is viewed approximately face-on.
The southern X-ray extraction includes most of 
the high surface brightness region measured in the UV continuum
\citep{Gon98}.
Edge-on starbursts, 
such as M82  and NGC 253, 
indicate the three-dimensional 
structure that the aperture covers  \citep{Str03}.
Specifically, the X-ray- and UV-brightest 
regions of intense star formation are located at the
base of an outflowing superwind, which extends conically out
of the plane of the galaxy.
The 0.7-keV thermal component is most similar to 
the diffuse outflow measured in the edge-on cases, where
it can be spatially isolated.  
The 200-pc radial scale of NGC 5135's bright starburst is comparable
to the intense cores of edge-on starbursts, so we expect that 
the outflow extends on similar physical scales---on the 
order of a kpc or two---along the line of sight.

The diffuse region D1 is spectrally similar to
the intense starburst, with the same hard power-law and soft thermal
components, although at greatly diminished surface brightness.
 These two regions may be genuinely
related, representing the outer walls of the expanding
conical outflow, which is projected onto the starburst base
at small scale height and has a larger projected radius
at large scale height.
The more extended component, D2, arises in the larger 
(4 kpc scale)
star-forming area that the optical and UV \hst{} images show, but 
X-ray emission does not cover the entire galaxy disk.
Thus, even this extended component is closely tied to
the sites of significant star formation, so this 
diffuse emission is
most likely a halo rather than disk component of NGC 5135's interstellar
medium. 

\subsection{Fluorescent Iron Emission\label{subsec:fek}}
The properties of the Fe line measured in the AGN spectrum
are consistent with
fluorescence in a cold, ``neutral'' medium.  The spectrally unresolved
line limits the temperature to $T < 2\times 10^6$ K.
The energy of Fe K$\alpha$ is about 6.4 keV for all ions
up to \ion{Fe}{17}, and only reaches 6.7 and 6.9 keV
for He-like and H-like Fe, respectively.
Thus, the Fe line we observe in NGC 5135 is not broad or
ionized enough to arise in a scattering medium that views 
the intrinsic continuum directly.
In these more sensitive and higher resolution data, 
we do not find any evidence for the strong ionized Fe lines reported by
\citet{Tur97} based on {\it ASCA} observations.  
The hard
X-ray fluxes of these two observations separated by over 5 years
are similar, so AGN variability is an unlikely explanation
for this discrepancy. 

The large fluorescent iron equivalent width, 
EW $ = 2.4$ keV,
further constrains the geometry of material close to the AGN.
In the simplest case, where both the
continuum and Fe lines are viewed directly, 
EW $\lesssim 250$ eV \citep{Kro87}, which is typical of Seyfert 1 galaxies
\citep[e.g., ][]{Nan94}. 
Reaching EW $ > 1$ keV requires high column density ($N_H > 10^{24} \psc$)
in the fluorescing region \citep{Ghi94,Kro94}.

\begin{figure}[thb]
\includegraphics[width=0.5\textwidth]{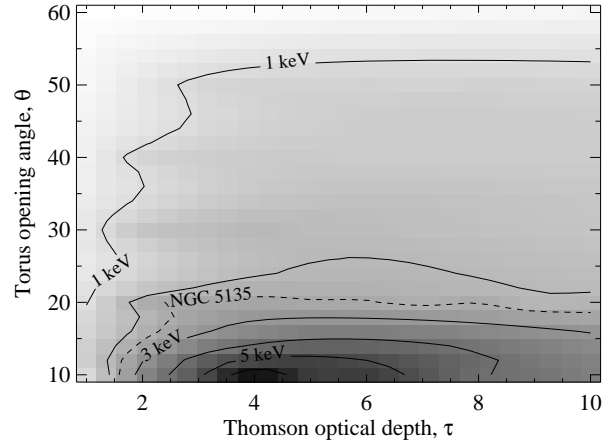} 
\caption{\label{fig:ewplot}
Fe K$\alpha$ line EW as a function of Thomson optical depth and
opening angle of the obscuring region \citep{Kro94,Lev02}.
The extremely large EW of NGC 5135 (dashed line) requires an
opening angle $\theta \le 20^\circ$, or equivalently a covering fraction
$\ge 90\%$.
}
\end{figure}

Detection of an extremely large EW strongly constrains the geometry
of the obscurer.  Extending the simulations of \citet{Kro94},
\citet{Lev02} show
that variations of geometry, not iron abundance or intrinsic spectral 
shape, are required to produce EWs significantly greater than 1 keV.
Specifically, the largest EWs require that the solid angle
covering fraction of the AGN is also large.  
Equivalently, 
we may describe this situation as a small opening angle of a
toroidal obscuring region.
In NGC 5135, for example, the measured EW constrains the 
opening angle, $\theta \le 20^\circ$,
or covering fraction $\ge 90\%$.  
Figure \ref{fig:ewplot} illustrates the effect of 
variations of $\theta$ and Thomson optical depth along the
line of sight, $\tau$, on the EW.  The Fe K$\alpha$ EW 
measured in NGC 5135 is also plotted.

The large covering fraction may be related to the circumnuclear starburst 
in this case.  The mechanical energy of stellar winds and supernovae
may help to ``puff'' the obscuring torus of material to become
more spherical than disk-like.  Whatever the origin, the large
covering fraction means that very few lines of sight to this
galaxy have a direct view of the central engine.  Galaxies 
similar to NGC 5135 are therefore extremely unlikely to be observed as
Seyfert 1s.  

The luminosity of the K$\alpha$ line is related to the intrinsic
luminosity of the AGN, although the uncertainty in the
exact correlation is large.
Based on the 
K$\alpha$ luminosity, we estimate the 
intrinsic 2--10 keV luminosity of the AGN
$L_{2-10, int} = 1 \times 10^{43} {\rm \, erg\, s^{-1}}$, 
considering opening angles $\theta \le 20^\circ$ in
the Monte Carlo simulations of \citet{Kro94} and \citet{Lev02}.
As expected in this Compton thick AGN, less than 1\%
of the 
intrinsic X-ray luminosity emerges from the obscured region 
at hard X-ray energies;
the observed 2--10 keV luminosity, 
$L_{2-10, AGN} = 3 \times 10^{40} {\rm \, erg\, s^{-1}}$.
Scaling the predicted intrinsic hard X-ray luminosity to
the observed spectral energy distributions of unobscured
quasars \citep{Elv94}, we estimate the bolometric luminosity
of the AGN below 10 keV; 
$L_{bol, AGN} = 3 \times 10^{44} {\rm \, erg\, s^{-1}}$,
which is only half of the bolometric luminosity
of the galaxy as a whole (\S \ref{sec:sed}).

\subsection{Absorption and Geometry\label{subsec:absorb}}
Along the direct line of sight to the AGN we
find $N_H > 10^{24} \psc$,
based on the large Fe K$\alpha$ EW and the absence
of a continuum characteristic of a directly-viewed
AGN, even at hard X-ray energies.
The optical [\ion{O}{3}] $\lambda 5007$ luminosity relative to 
the 2--10 keV emission offers
further support for the Compton thick classification
of NGC 5135.
The [\ion{O}{3}]  emission is expected to arise
outside the obscuring region, 
while the emergent X-ray continuum
may be suppressed along the line of sight.
A ratio $L_{2-10}/L_{[O\ III]} < 1$ is characteristic of
Compton thick cases \citep{Mai98}.
In a 2--4$\arcsec$ aperture, \citet{Whi92} measures a flux 
$f_{[O\ III]} = 2.2\times10^{-13} {\rm \, erg\, cm^{-2}\, s^{-1}}$.
Correcting for reddening  based on the Balmer decrement 
\citep{Gon98}, we find 
$L_{2-10}/L_{[O\ III]} = 0.08$.
Starburst emission certainly  contributes to the measured
[\ion{O}{3}] luminosity, but this result 
is at least consistent
with Compton thick obscuration of the AGN.

The starburst itself is responsible for some of
this obscuration.
In general, the circumnuclear starburst requires
a concentration of mass, some of which is located
in front of the active nucleus.  In this case, we use
the star formation rate measured within the
$1\farcs7 \times 1\farcs7$ aperture
of the Goddard High Resolution Spectrograph (GHRS) on {\it HST}
to estimate the column density near the nucleus.
A  star-formation rate 
$SFR \approx 2 M_\sun {\rm \, yr^{-1}}$ \citep{Gon98}
within this aperture corresponds to a mean gas column density
$N_H \approx 10^{23}\psc$, using the conversion
of \citet{Ken98}.
In extreme star-forming galaxies, such as Arp 220,
the observed $SFR$ alone produces Compton thick column
densities along the line of sight.

We find slightly lower column densities in the \chandra{}
observations of the starburst areas, including
the soft thermal emission in the North region, because
we do not directly probe the line of sight to the very center
of NGC 5135.
 We do, however, always measure  absorption
in excess of the Galactic column density  alone.
The X-ray measured obscuration, $N_H = 4\times10^{21}\psc$, 
is greater than that measured within the starburst itself
in the UV, where 
$E(B-V) = 0.20$, or $N_H = 1.2\times 10^{21}\psc$
in the comparable GHRS aperture.
The UV extinction relies on modelling the
observed emission features assuming a foreground screen, 
and therefore represents only a lower limit on the total extinction
through the entire starburst region.  Sources located
at $\tau \lesssim 1$ dominate the observed emission.
Thus, it is not surprising that the detected AGN, located
behind the entire obscuring starburst, is even more absorbed than
the thermal emission alone,
or that the UV measurement  yields a smaller column density.
The UV-determined extinction is comparable to the obscuration measured in
the X-ray spectrum of region D1.
The column density of the D2 extraction is greater.  We note that
in other UV spectra,  \citet{Gon98} also measure
greater obscuration  in larger apertures in NGC 5135 and other
Seyfert 2/starburst galaxies.

\begin{deluxetable}{lrr}
\tablewidth{0pt}
\tablecaption{Region Luminosities\label{tab:lum}}
\tablehead{
\colhead{Region}
&\colhead{$L_{0.5-2}$}
&\colhead{$L_{2-10}$}\\
&\colhead{($10^{40}{\rm \,erg\,s^{-1}}$)}
&\colhead{($10^{40}{\rm \,erg\,s^{-1}}$)}
}
\startdata
NGC 5135 (total)&  18\phantom{.2}  &    19\phantom{.2} \\
North (AGN)     &  2.6 &   9.4 \\
South  (SB)     &  4.6 &   1.5 \\
D1              &  3.1 &   1.6 \\
D2              &  1.5 &   0.5 \\
\enddata

\end{deluxetable}

\section{Far-Infrared Relations and Spectral Energy Distributions}\label{sec:sed}
NGC 5135 is classified as a normal Seyfert 2 galaxy on the basis of optical
emission-line ratios, yet at most wavelengths, the AGN
contributes only a small fraction of the detectable luminosity. 
Even in X-rays, the starburst is significant.
Table \ref{tab:lum} lists the soft and hard X-ray luminosities of the four 
spectroscopic regions and 
a 1$\arcmin$ aperture, which represents the total emission of NGC 5135.

The far-infrared (FIR) emission of NGC 5135 is characteristic of star-forming
galaxies, both in spectral shape and relative magnitude.  While active
galaxies typically exhibit 25- to 60-$\mu$m flux ratios
$f_{25}/f_{60} > 0.26$ \citep{deG87} in the arcminute-scale apertures
of the {\it Infrared Astronomical Satellite} ({\it IRAS}), 
$F_{25}/F_{60}=0.14$ in NGC 5135.
The FIR luminosity $L_{FIR} = 5.5\times10^{44} {\rm \, erg\, s^{-1}}$,
and the total 8--1000$\mu$m luminosity
$L_{IR} = 7.1\times10^{44} {\rm \, erg\, s^{-1}}$, following
the prescriptions of \citet{Hel85} and \citet{San96} to calculate
these luminosities from {\it IRAS} fluxes.
For starburst galaxies, $L_{bol} \approx L_{IR}$, and in this
case, the observed IR luminosity is  more than twice the predicted 
intrinsic luminosity of the buried AGN.

The bolometric luminosity is related to the star formation rate, with
$SFR = L_{bol}/2.8\times10^{43}$ \citep{Lei95}.  Here we
find $SFR = 25 M_{\sun}\, {\rm yr^{-1}}$.
As expected, this is greater than the UV $SFR$ measured within the
central 500-pc GHRS aperture.
It is less than the extinction-corrected 
$SFR \approx 50 M_{\sun}\, {\rm yr^{-1}}$ derived from the 
{\it International Ultraviolet Explorer} ({\it IUE} )
data \citep{Gon98}, but because the correction for 
extinction is highly uncertain and large, UV- and IR-measured
$SFR$s often disagree \citep{Meu99}.

The ratios of total X-ray to FIR luminosity measured in NGC 5135
are characteristic of starburst
galaxies, not AGN.  In Seyfert 1 and Seyfert 2 galaxies that
do not contain starbursts, $L_{2-10}/L_{FIR} \ge 10^{-2}$,
while this ratio tends to be much lower in
starburst and ultraluminous infrared galaxies \citep{LWH,Pta03}.
We find $L_{2-10, total}/L_{FIR} = 4.8\times10^{-4}$ in
NGC 5135, where $L_{2-10, total}$ is the total observed
2--10 keV luminosity.
Considering only the non-AGN contribution, which we take to be
the North region contribution subtracted from the total,
$L_{2-10, SB}/L_{FIR} = 2.4\times10^{-4}$, exactly the linear
relationship \citet*{Ran03} fit to a local sample of starburst
galaxies.  
The soft X-ray/FIR ratios of NGC 5135 
($L_{0.5-2, total}/L_{FIR} = 4.6\times10^{-4}$ and
$L_{0.5-2, SB}/L_{FIR} = 3.9\times10^{-4}$)
are slightly higher than the starburst average, 
but within the typical scatter of the data.

\begin{figure}[htb]
\includegraphics[width=0.5\textwidth]{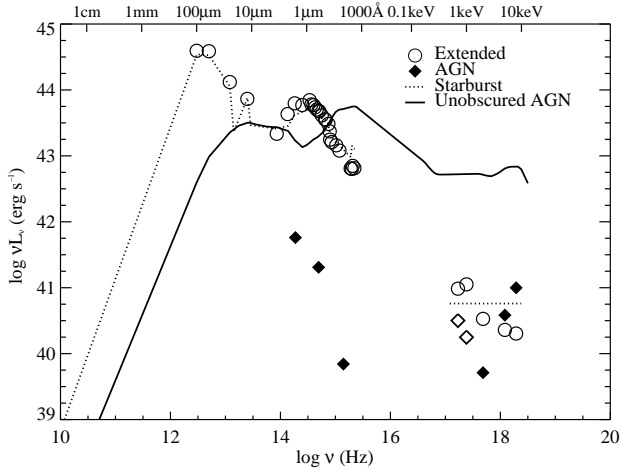} 
\caption{\label{fig:sed}
Spectral energy distribution of NGC 5135, showing large-scale emission
(open circles) and the spatially isolated AGN  contributions (filled diamonds).
Thermal contributions to the nuclear X-ray spectrum are
marked separately (open diamonds).
In general, the galaxy appears very similar to dusty
starburst galaxies (dotted line; \citealt{Schm97}).  
For comparison, the average radio-quiet quasar spectrum \citep{Elv94} scaled to
the intrinsic bolometric luminosity of NGC 5135 is shown 
(solid line).  Simple foreground extinction of this spectrum
cannot reproduce the unresolved emission. 
The emergent X-ray spectrum is strongly absorbed and reprocessed,
and the SED near optical wavelengths indicates the presence of
some additional foreground extinction.
}
\end{figure}

Figure \ref{fig:sed} illustrates the spectral energy distributions (SEDs) of
extended emission and the AGN separately. 
Across the spectrum, NGC 5135 as a whole 
(plotted with open circles) 
appears most similar to dusty starburst
galaxies, whose average SED \citep{Schm97} is plotted, 
scaled to the luminosity of NGC 5135.
Here we estimate the average X-ray SED
from the FIR measurements using the relationship of \citet{Ran03}.
The large-aperture data are compiled from the {\it IRAS} 
Faint Source Catalog, ground-based observations of \citet{Gla85},
and {\it IUE} measurements of \citet{Sto95}.
The ``extended'' X-ray data that are indicated 
at low spectral resolution 
come from the measurements of the entire galaxy {\em excluding} the
central $2\farcs5$.  (This excluded region, slightly larger than
the AGN aperture, was selected  to avoid all AGN contamination.)
The solid filled X-ray points are measured in the North region
and include only the spectral components characteristic of the AGN.
The data points are derived from monochromatic fluxes.   The
highest energy data plotted (at 5 and 8 keV) indicate
continuum flux only and exclude the Fe K$\alpha$ line.
The open diamonds represent the thermal emission measured in the
nuclear region, which we do not attribute immediately to the AGN.

In the UV, optical, and NIR, we measure the AGN alone 
by fitting the appropriate two-dimensional point-spread function (PSF) to 
images from the \hst{} observations mentioned above.
We modeled the optical and NIR PSFs with 
Tiny Tim\footnote{http://www.stsci.edu/software/tinytim} and
used the observation of a point source for an empirical UV
PSF.  The formally best-fitting  PSF subtraction
within a limited area produces unphysical residuals that
decline toward the center 
and therefore yields an upper limit on the unresolved
emission \citep{Rid97}.
We constrain the residuals 
(which represent genuinely extended emission)
to be monotonic across small interior regions for a more
realistic estimate of the AGN contribution.
This constraint reduces the AGN flux by about
10\% in the NIR and 25\% in the optical. It is
most restrictive in the UV, where the truly
unresolved
emission appears to be only one-third of the 
flux within the central region.
For the unresolved emission, we find 
$6.7\times 10^{-3}$, $9.0 \times 10^{-5}$,  and
$1.1\times10^{-6}$ Jy in the NIR, optical, and 
UV, respectively.

Where the AGN can be spatially isolated, its
spectral energy distribution  does not match 
that of observed radio-quiet quasars \citep{Elv94}.
A simple screen of obscuration 
in the foreground of the average unobscured quasar also fails to describe the
AGN SED.  The Compton thick reprocessing, which strongly alters the X-ray spectrum,
is essential.  Although spatially unresolved, the observed lower-energy emission 
must also be reflected; 
we would not detect the AGN directly at these wavelengths behind a
column density of $10^{24}\psc$.  The steep decline of
the relative luminosity from the
NIR through UV, however,  indicates that
some additional foreground obscuration is also present.

This example illustrates the difficulty of identifying an AGN in the presence
of a starburst, especially more distant cases where spectral apertures 
cover larger physical scales and signal-to-noise ratios are diminished.
The characteristic optical
emission lines are easily diluted when off-nuclear
sources account for a larger fraction of the light \citep{Mor02,Kau03}, or
when a strong starburst is also present \citep{Lev01}.

Only above 4 keV does the AGN dominate the detectable emission from NGC 5135, 
yet simple modeling of the total X-ray spectrum
would yield misleading results.  
The detected 2--10 keV luminosity could be produced by a starburst alone.
The spatially-integrated spectrum is relatively flat from 0.4--8 keV,
offering no indication of the distinct physical origin of the
soft and hard components.
The strong obscuration of the AGN 
is not at all evident  in the
spatially-integrated spectrum.
A crude  technique such as spectral hardness ratios would incorrectly
suggest the presence of a weak yet unobscured AGN.
The example of NGC 5135 demonstrates how such 
Compton thick AGN may therefore remain successfully hidden at most wavelengths
where surveys have been performed, particularly when they
are located at large redshift.

Compton thick examples are truly normal AGN,
comprising roughly half of samples that are not selected on
the basis of their X-ray emission \citep{Ris99}.
They do not necessarily appear to be X-ray bright below 10 keV.
Stellar processes further complicate their soft X-ray spectra,
since the large reservoirs of gas that block the AGN can also
form stars, which subsequently produce soft thermal emission.  

Though difficult to identify, such obscured sources may contribute
significantly to the cosmic X-ray background.
Synthesis models that combine a distribution in both
absorption to the central engine and redshift \citep{Set89,Mad94,Com95}
are essential to reproduce the background spectrum, which peaks around 30 keV.
While discrete sources account for 80\% or more of the XRB
at energies up to 10 keV \citep{Bra01,Gia01,Has01},
these observed redshift and column density distributions do not 
spectrally match  the X-ray background  
at higher energies \citep{Gil03}.
Furthermore, theoretical models  \citep{Fab98} suggest 
that up to 90\% of the accretion power in the Universe is hidden
from direct view.  NGC 5135 offers a local demonstration
of how such sources may remain effectively buried behind large
column densities and along most lines of sight.

\section{Conclusions} 
In analyzing this \chandra{} ACIS observation of the Seyfert 2/starburst
galaxy NGC 5135, we spatially and spectrally distinguish both the
AGN and the stellar origin of its X-ray emission.
Images reveal two strong and concentrated central sources.  The northern of these
is the AGN, while the southern is associated with star formation and 
likely represents the base of an outflowing starburst superwind.
Recognizing the contributions of both the AGN and stellar
processes, we apply physically-motivated models to spectroscopy
of several distinct regions of interest.  
In addition
to the two central sources, these regions include two extended
areas, which likely represent a hot halo component of NGC 5135.
In the nuclear region, the AGN produces
 a very flat ($\Gamma=0$) power law
continuum and extremely prominent (EW $=2.4$ keV) Fe K$\alpha$
fluorescence line.  These spectral characteristics demonstrate
that the central engine is highly obscured,
behind $N_H > 10^{24} \psc$.  The intrinsic emission emerges
only after being strongly reprocessed and is diminished by
factors of 100 as observed in the \chandra{} bandpass.
This central region encompasses a physical scale of 340 pc,
which includes sites of prominent star formation that are viewed
directly in UV through NIR wavelengths.  These stars are responsible
for the soft thermal emission we measure in this region.

The remainder of the X-ray emission, including the southern source,
is spatially resolved.  We attribute this emission to star
formation. 
Prominent soft thermal X-rays characterize these spectra.  They
also require some additional continuum components, which 
are responsible for their higher-energy ($E > 4$ keV)
luminosity.  Unresolved populations of X-ray binaries  likely
produce this emission.

The vast majority of the detected soft X-ray luminosity from NGC 5135
($L_{0.5-2} = 1.8\times10^{41} {\rm \, erg\, s^{-1}}$) is spatially
extended.  The southern source accounts for fully one-quarter of this
emission.  The AGN itself accounts for almost none of it, given
that nearly all of the soft X-rays within the nuclear aperture are
thermal.  The AGN is directly responsible for roughly half the
detected hard X-ray luminosity 
($L_{2-10} = 1.9\times10^{41} {\rm \, erg\, s^{-1}}$).
Based on the luminosity of the Fe K$\alpha$ line, we
estimate the intrinsic bolometric luminosity 
of the AGN:
$L_{bol, AGN} = 3 \times 10^{44} {\rm \, erg\, s^{-1}}$.
This AGN luminosity is
only half of the bolometric luminosity
of the galaxy as a whole,
$L_{bol} \approx 7 \times10^{44} {\rm \, erg\, s^{-1}}$, 
based on the IR luminosity.

We suggest that the starburst is related to the obscuration
of the AGN.  The starburst itself directly 
hides the AGN, at least in part.
The measured star formation rate alone can account for 
10\% of the indicated absorption.  
A starburst requires large reservoirs of gas in the
center of the galaxy, and the
dynamic conditions that concentrate this material 
also serve to further obscure the AGN.  
The starburst
may also be related to the {\em activity} of the central engine,
with  the instabilities that lead to star
formation aiding accretion.

Although optically classified as an ordinary Seyfert 2 galaxy,
the Fe K$\alpha$ emission is the only X-ray feature
that truly identifies the presence of an AGN.  The extremely
large EW of the line requires a very large (90\%) covering fraction
of material, allowing few direct views of the central
engine from any line of sight.
The Compton thick obscuration severely diminishes the
hard X-ray emission that is a useful discriminant
of less-obscured AGN.

NGC 5135 represents a local case study of some of the
complex reality that is also likely present in more distant
examples.  
Even in X-rays, stellar processes are significant, and
combined with 
Compton thick obscuration, a deeply buried AGN will not
be obvious in data that have
low effective spatial or spectral resolution.
These are particular difficulties in the study of
ultraluminous infrared galaxies, in which star formation
is known to be important, and also in the resolution
of the X-ray background, where large
obscuration along most lines of sight
diminishes the likelihood of identifying the hidden AGN
that are essential to produce the background spectrum
at higher energies.

\acknowledgements

We thank B. McKernan for providing XSTAR photoionization models,
A. Kinkhabwala for the observed spectrum of NGC 1068,
and an anonymous referee for useful suggestions.
We thank D. K. Strickland for advice on identifying sources
and S. E. Ridgway for many discussions about PSF subtraction.
This research has made use of the NASA/IPAC Extragalactic Database
(NED) which is operated by the Jet Propulsion Laboratory, California
Institute of Technology, under contract with the National Aeronautics
and Space Administration.
The Digitized Sky Survey was produced at the Space Telescope Science
Institute under U.S. Government grant NAG W-2166. The Second Epoch image
presented here 
was made by the Anglo-Australian Observatory with the UK Schmidt Telescope.
Part of this work was based on observations made with the NASA/ESA Hubble
Space Telescope, obtained from the data archive at the Space Telescope
Science Institute. STScI is operated by the Association of
Universities for Research in Astronomy, Inc. under NASA contract NAS
5-26555.
This work is supported by NASA grant G01-2119.

\end{document}